\documentclass[useAMS,usenatbib]{mn2e}
\usepackage[dvips]{graphicx}
\title[Monthly Notices: \LaTeXe\ guide for authors]
  {Morphology Classification and Photometric Redshift Measurement of Galaxies}
\author[Y. Zhang et~al.]
  {Yanxia Zhang$^{1}$\thanks{Email: zyx@lamost.org, yzhao@lamost.org}
  ,Lili Li$^{1,2}$
  and Yongheng Zhao$^1$ \\
  $^1$National Astronomical Observatories,
Chinese Academy of Sciences, 20A Datun Road, Chaoyang District,
100012, Beijing, P.R.China\\
  $^2$Weishanlu Middle School, 300222, Tianjin, P.R.China}
\date{Released 2002 Xxxxx XX}

\pagerange{\pageref{firstpage}--\pageref{lastpage}} \pubyear{2002}

\def\LaTeX{L\kern-.36em\raise.3ex\hbox{a}\kern-.15em
    T\kern-.1667em\lower.7ex\hbox{E}\kern-.125emX}

\begin{document}

\label{firstpage}

\maketitle

\begin{abstract}

Based on the Sloan Digital Sky Survey Data Release 5 Galaxy Sample,
we explore photometric morphology classification and redshift
estimation of galaxies using photometric data and known
spectroscopic redshifts. An unsupervised method, k-means algorithm,
is used to separate the whole galaxy sample into early- and
late-type galaxies. Then we investigate the photometric redshift
measurement with different input patterns by means of artificial
neural networks (ANNs) for the total sample and the two subsamples.
The experimental result indicates that ANNs show better performance
when the more parameters are applied in the training set, and the
mixed accuracy $\sigma_{\rm mix}=\sqrt{{\sigma_{\rm
early}}^2+{\sigma_{\rm late}}^2}$ of photometric redshift estimation
for the two subsets is superior to $\sigma_{\rm z}$ for the overall
sample alone. For the optimal result, the rms deviation of
photometric redshifts for the mixed sample amounts to 0.0192, that
for the overall sample is 0.0196, meanwhile, that for early- and
late-type galaxies adds up to 0.0164 and 0.0217, respectively.

\end{abstract}

\begin{keywords}
catalogs -¡ª galaxies: distances and redshifts -¡ª galaxies: general
-¡ª galaxies: photometry -¡ª surveys ¡ª- techniques: photometric
\end{keywords}

\section{Introduction}

The Sloan Digital Sky Survey (SDSS, York et~al. 2000) is an
astronomical survey project, which covers more than a quarter of the
sky, to construct the first comprehensive digital map of the
universe in 3D. The large amount of spectroscopic and photometric
data obtained during the last years by SDSS, which has opened a new
horizon for the study of galaxy properties such as galaxy evolution,
clusters, redshifts, large-scale distribution on morphological type
and so on. However, photometric classification and redshift
estimation is of prime importance for the SDSS project. Obtaining
reliable object type and redshift estimation based on SDSS
photometry is thus an extremely valuable adjunct to the
spectroscopic sample.

One of the first segregation discovered in galaxy clusters was the
morphological one. The first evidences of such segregation date from
Curtis (1918) and Hubble \& Humason (1931), and was quantified by
Oemler (1974) and Melnick \& Sargent (1977). The problem of general
automated classification always lies in the difficulty of finding
quantitative measures that strongly correlate with the Hubble
sequence based on visual inspections. Shimasaku et~al. (2001) and
Strateva et~al. (2001) using SDSS data, showed that the ratio of
Petrosian 50 percent light radius to Petrosian 90 percent light
radius, $C_{in}$, measured in the $r$-band image was a useful index
for quantifying galaxy morphology. For early-type galaxies,
concentration index $C_{in}$ is larger than 2.5; while for late-type
galaxies,  $C_{in}$ is less than 2.5. Strateva et~al. (2001) also
found that the color $u-r=-2.22$ efficiently separates early- and
late-type galaxies at $z<0.4$. The basis for the classification of
the SDSS photometric database can be provided by the objects whose
nature is precisely known from spectroscopy.

Photometric redshifts refer to the redshift estimation of galaxies
using only medium- or broad-band photometry or imaging instead of
spectroscopy. Techniques for deriving redshifts from broadband
photometry were pioneered by Baum (1962). Subsequent implementations
of these basic techniques have been made by Couch et~al. (1983) and
Koo (1985). In terms of data mining, the photometric redshift
estimation belongs to the regression task of data mining. In
principal, the various approaches used for solving regression
problem may be applied to the photometric redshift measurement. So
far there has been a great amount of research on the techniques of
photometric redshift estimation. The techniques are broadly grouped
into three kinds: the template-matching method, the empirical
training-set method, instance-based learning method. When using the
template-matching method, we must have the template. The quality of
template directly influences the performance of predicting
photometric redshifts. The template spectra come from population
synthesis models (e.g. Bruzual \& Charlot 1993) or from spectra of
real objects (e.g. Coleman et al. 1980). The empirical training-set
method is based on the real data. So whether the real data is enough
and complete is an important factor. The training-set method is
usually implemented by train-test method or cross-validation method,
in other words, it needs to train training-set to get a classifier
or regressor and then the classifier or regressor is tested by test
set. Typical empirical training-set methods include artificial
neural networks (ANNs, Collister \& Lahav 2004; Firth, Lahav \&
Somerville 2003; Vanzella et~al. 2004; Li et~al. 2007), support
vector machines (SVMs, Wadadekar 2005; Wang et~al. 2007, 2008),
ensemble learning and Gaussian process regression (Way \& Srivastava
2006), and linear and non-linear polynomial fitting (Brunner et~al.
1997; Wang, Bahcall \& Turner 1998; Budav$\acute{a}$ri et al. 2005;
Hsieh et~al. 2005; Connolly et~al. 1995). Although the
instance-based learning method also relies on the real data, it is
different from the training-set method for it has no training
process and stores all data in the memory of computer. Examples of
such techniques are k-nearest neighbours (e.g. Csabai et~al. 2003;
Ball et~al. 2007; Gao, Zhang \& Zhao 2007), kernel regression (Wang
et~al. 2007, 2008), and locally weighted regression.

This paper majors in morphological classification of galaxies using
k-means algorithm and photometric redshift estimation of galaxies
using artificial neural networks (ANNs). The paper is organized as
follows. Section 2 gives the scheme of this paper. Section 3
introduces a brief overview of k-means algorithm and artificial
neural networks, respectively. Section 4 describes photometric
classification of galaxies using k-means algorithm. We investigate
redshift estimation using an extensive series of tests in Section 5.
The conclusions and discussions are summarized in Section 6.

\section{The Scheme of This Paper}

This paper demonstrates the potential of bulk classification of the
SDSS data and indicates a wide range of research applications,
especially for redshift estimation. K-means algorithm offers an
efficient way to identify the physical nature of SDSS sources, so it
has a strong potential to become an important classification tool
for the bulk of the SDSS photometric database. We collect the SDSS
Data Release 5 galaxy sample. Then k-means algorithm is applied on
this sample for two respects: one is to preprocess the sample by
removing outliers; another is to automatically separate preprocessed
sample into two morphological classes (namely early- and late-type
galaxies). After that, we consider two cases for photometric
redshift estimation with different input patterns by artificial
neural networks (ANNs). The first is directly to use ANNs on the
total preprocessed sample, and the other is to employ ANNs on the
early- and late-type galaxies respectively. Finally the results of
the two cases are compared.

\section{Principle}
\subsection{K-means Algorithm}

The k-means algorithm (MacQueen, 1967) is one of the simplest
unsupervised learning algorithms used for clustering problem. The
algorithm clusters $n$ objects based on attributes into $k$
partitions, $k<n$. The main idea is to define $k$ centroids, one
for each cluster. These centroids should be placed in a cunning
way because different locations cause different results. So, the
better choice is to place them as much as possible far away from
each other. The next step is to take each point belonging to a
given data set and associate it to the nearest centroid. When no
point is pending, the first step is completed and an early
groupage is done. At this point we need to re-calculate $k$ new
centroids as barycenters of the clusters resulting from the
previous step. After we have these $k$ new centroids, a new
binding has to be done between the same data set points and the
nearest new centroid. A loop has been generated. As a result of
this loop we may notice that the $k$ centroids change their
location step by step until no more changes are done. In other
words centroids do not move any more.

K-means algorithm is similar to the expectation-maximization
algorithm for mixtures of Gaussians in that they both attempt to
find the centers of natural clusters in the data. It assumes that
the object attributes form a vector space. The objective it tries to
achieve is to minimize total intra-cluster variance, or, the squared
error function
\begin{equation}
V=\sum\limits_{i=1}^{k}\sum\limits_{x_j\in S_i}(x_j-\mu_i)^2
\end{equation}
where there are $k$ clusters $S_i$, $i=1,2,3,...,k$,  and $\mu_i$ is
the centroid or mean point of all the points $x_j\in S_i$.

Although it can be proved that the procedure will always terminate,
the k-means algorithm does not necessarily find the most optimal
configuration, corresponding to the global objective function
minimum. The algorithm is also significantly sensitive to the
initial randomly selected cluster centers. The k-means algorithm can
be run multiple times to reduce this effect.

\subsection{Artificial Neural Networks}

An Artificial Neural Network (ANN) is an information processing
paradigm that is inspired by the way biological nervous systems,
such as the brain, process information. The key element of this
paradigm is the novel structure of the information processing
system. ANNs are collections of interconnected neurons each
capable of carrying out simple processing. Thus, they are composed
of massively parallel distributed processors that have an inherent
property of storing experiential knowledge and making it available
for use. The knowledge is acquired by the network through a
learning process and is stored in interneuron connection strengths
- known as synaptic weights (Haykin 1994). Practical applications
of ANNs most often employ supervised learning. For supervised
learning, one must provide training data that includes both the
input (a set of vectors of parameters, here each vector
corresponds to a galaxy) and the desired result or the target
value (the corresponding redshifts). After the network is trained
successfully, one can present input data alone to the ANN (that
is, input data without the desired result), and the ANN will
compute an output value that approximates the desired result. This
is achieved by using a training algorithm to minimize the cost
function which represents the difference (error) between the
actual and desired output. The cost function $E$ is commonly of
the form
\begin{equation}
E=\frac{1}{p}\sum^{p}_{k=1}(o_k-t_k)^2,
\end{equation}
where $o_k$ and $t_k$ are the output and target respectively for the
objects, $p$ is the sample size. Generally the topology of an ANN
can be schematized as a set of $N$ layers (see Fig.~1), with each
layer composed of a number of neurons. The first layer ($i=1$) is
usually called the ``input layer", the intermediate ones the
``hidden layers", and the last one ($i=N$) the ``output layer". Such
a species of ANN is formally known as a ``multilayer perceptron"
(MLP). Each neuron $j$ in the $s$ layer derives a weighted sum of
the $M$ output $z^{s-1}_i$ from the previous layer ($s-1$) and,
through either a linear or a non-linear function, produces an
output,
\begin{equation}
z^{(s)}_i=f(\sum^M_{i=0}w^{(s)}_{ji}z^{(s-1)}_i).
\end{equation}
Here $w_{j0}$ denotes the bias for the hidden unit $j$, and $f$ is
an activation function such as the continuous sigmoid or, as used
here, the tanh function, which has an output range of -1 to 1:
\begin{equation}
f(x)=\frac{2}{1+e^{-2x}}-1.
\end{equation}
When the entire network has been executed, the output of the last
layer is taken as the output of the entire network. The free
parameters of ANNs are the weight vectors. During the training
session, the weights of the connections are adjusted so as to
minimize the total error function. The learning procedure is the
so-called ``back propagation". The number of layers, the number of
neurons in each layer, and the functions are chosen from the
beginning and specify the so called ``architecture" of the ANN.
Neural networks learn by examples. The neural network user gathers
representative data into a training set and initiates the weight
vector with a random seed, then invokes the training algorithms to
automatically learn the structure of the data. Here, we use a method
that is popular in neural network research: the Levenberg-Marquardt
method (Levenberg 1944; Marquardt 1963; also detailed in Bishop
1995). This has the advantage that it converges very quickly to a
minimum of the error function. This error function may not have just
a global minimum in the multidimensional weight space but could have
a number of local minima instead. In general, network trained using
exactly the same training set for the same given number of epochs
but using different initial weights ( different starting points in
this space) will converge to slightly different final weights. In
order to avoid (possible) over-fitting during the training, another
part of the data can be reserved as a validation set (independent
both of the training and test sets, so not used in the updating of
the weights), and used during the training to monitor the
generalization error. After a chosen number of training iterations,
the training terminates and the final weights chosen for the ANN are
those from the iteration at which the cost function is minimal on
the validation set. This is useful to avoid over-fitting to the
training set when the training set is small, but the disadvantage of
this technique is that it reduces the amount of data available for
both training and validation, which is particularly undesirable if
the data set is small to begin with.

     \begin{figure}
   \begin{center}
   \includegraphics[bb=1 1 535 241,width=8cm,clip]{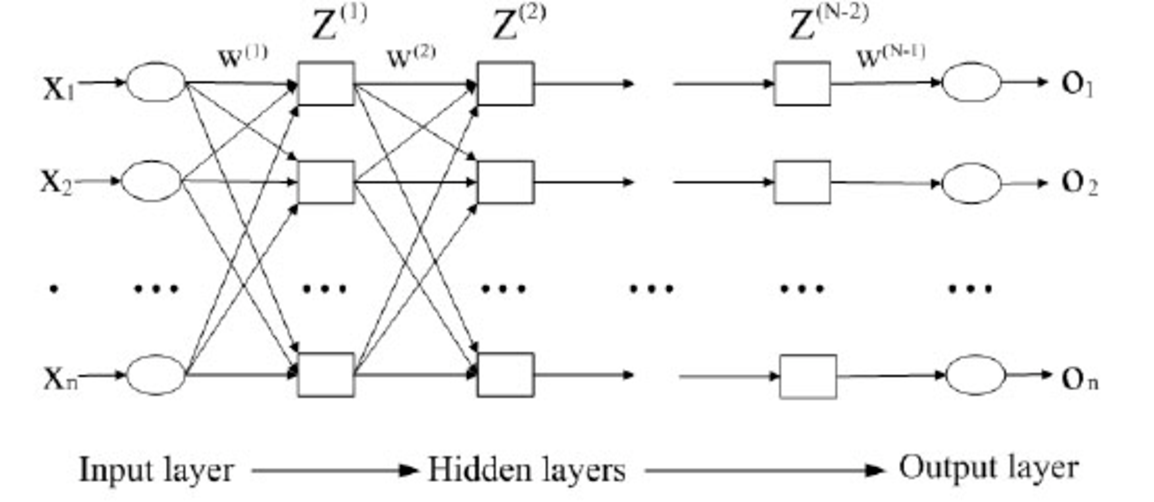}
   \end{center}
   \caption
   { \label{fig:example}
Schematic diagram of an ANN}
   \end{figure}

An ANN is configured for a specific application, such as pattern
recognition or data classification, through a learning process. ANNs
have various popular applications in astronomy, for example,
star/galaxy separation (e.g. Odewahn \& Nielsen 1994; Bertin \&
Arnouts 1996), morphological classification of galaxies (Nielsen \&
Odewahn 1994; Lahav et~al. 1996; Ball et~al. 2004), spectral
classification (Folkes et~al. 1996; Weaver 2000) and astronomical
objects classification (Zhang \& Zhao 2004, 2007), photometric
redshift estimation (e.g., Firth et~al. 2003; Vanzella et~al. 2004;
Li et~al. 2007; D'Abrusco et~al. 2007). As for a review of ANNs
applied in astronomy, refer to Serra-Ricart et~al. (1993), Miller
(1993), Storrie-Lombardi \& Lahav (1994) and Li et~al. (2006).
Bailer-Jones (1996, 2000) also majored in this issue.

\section{Morphology Classification}

\subsection{Chosen Galaxy Sample}

The Sloan Digital Sky Survey (SDSS) is the most ambitious
astronomical survey ever undertaken. The SDSS uses a dedicated,
2.5-meter telescope on Apache Point, New Mexico, equipped with two
powerful special-purpose instruments. The SDSS completed its first
phase of operations --  SDSS-I -- in June, 2005. Over the course of
five years, SDSS-I imaged more than 8,000 square degrees of the sky
in five bandpasses, detecting nearly 200 million celestial objects,
and it measured spectra of more than 675,000 galaxies, 90,000
quasars, and 185,000 stars. These data have supported studies
ranging from asteroids and nearby stars to the large scale structure
of the Universe. The SDSS has entered a new phase, SDSS-II,
continuing through June, 2008. SDSS-II will carry out three distinct
surveys ¡ª- the Sloan Legacy Survey, SEGUE, and the Sloan Supernova
Survey ¡ª- to address fundamental questions about the nature of the
Universe, the origin of galaxies and quasars, and the formation and
evolution of our own Galaxy, the Milky Way.

We downloaded 582,512 galaxies from the SDSS DR5 database, only
took objects with available five-band photometries. By removing
the records with default values, we obtained 582,257 galaxies.

\subsection{K-means Algorithm for Morphology Classification}

The origin of the morphology of galaxies is a longstanding issue
that could provide a key to discerning among models of the formation
of galaxies. Perhaps there is a general correlation between galaxy
color and Hubble morphologies, Strateva et~al. (2001) had
demonstrated that the galaxy color $u-r$ is related with the
morphology of galaxies and grouped galaxies into two families. In
this section we also used the galaxy color index to classify the
galaxies with k-means algorithm.

As we known, k-means algorithm is an unsupervised approach, and it
can automatically cluster based on the intrinsic property of
objects. Here we use this approach to cluster the above given sample
into two galaxy types (namely early-type and late-type galaxies).
For this experiment, we used the five Petrosian color index (u-g,
g-r, r-i, i-z, u-r) as the input parameters of k-means approach. The
algorithm programme may analyze the color property of the whole
sample and separate the given database into two classes
automatically. As a result, the number of each family is 300,903 and
281,354, respectively.

In order to verify the type of each class, we randomly select 1000
records from each family respectively to confirm their types, which
is achieved by using their position corresponding to the given one
in the NASA/IPAC Extragalactic Database (NED). By consulting, we
found that the cluster membership with 300,903 records are
early-type galaxies and the other 281,354 records belong to
late-type galaxies. In Fig.~2 and Fig.~3, we give the $u-r$
histogram for individual subclass (early- and late-type galaxies),
respectively. The $u-r$ histogram of the total galaxy sample is
shown in Fig.~4. The $g-r$ versus $u-r$ diagram is displayed in
Fig.~5, where the samples with red points are early-type galaxies
and the ones with black points are late-type galaxies. In Fig.~5,
the line is the $u-r=2.22$ plane. According to the $u-r$ cut
(Strateva et~al. 2001 ), the galaxies with $u-r>2.22$ belong to
early-type galaxies, while those with $u-r<2.22$ belong to late-type
galaxies. However, by means of k-means method for classification,
the $u-r$ value for early-type galaxies lies in the range from 2.0
to 8.0, that for late-type ones in the range from 0 to 5.0.
Therefore the classification results by the two methods show
difference. Only by the $u-r$ cut for morphological separation could
there be degeneracies. For example, some low redshift dusty edge-on
spirals could easily be misclassified as early-type galaxies. As a
result, the classification result of k-means method is more
reasonable than the $u-r$ cut because this algorithm applies more
information and classify objects in a multi-parameter space. K-means
algorithm has obvious strength that it is automatically clustering
by the properties of galaxies in the real universe and require no
additional assumptions about their formation and evolution.

     \begin{figure}
   \begin{center}
   \includegraphics[bb=32 179 346 402,width=8cm,clip]{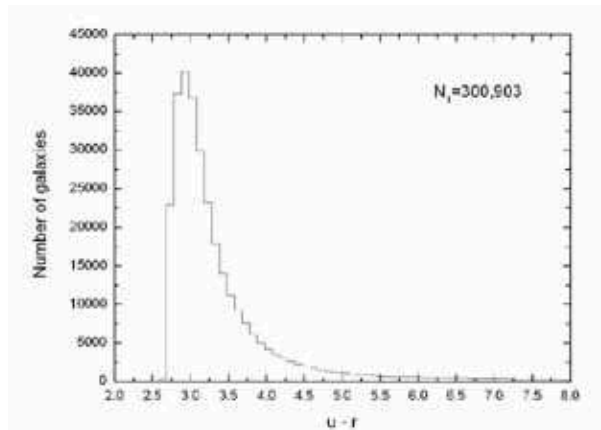}
   \end{center}
   \caption
   { \label{fig:example}
The $u-r$ histogram for early-type galaxies.}
   \end{figure}

     \begin{figure}
   \begin{center}
   \includegraphics[bb=32 179 346 402,width=8cm,clip]{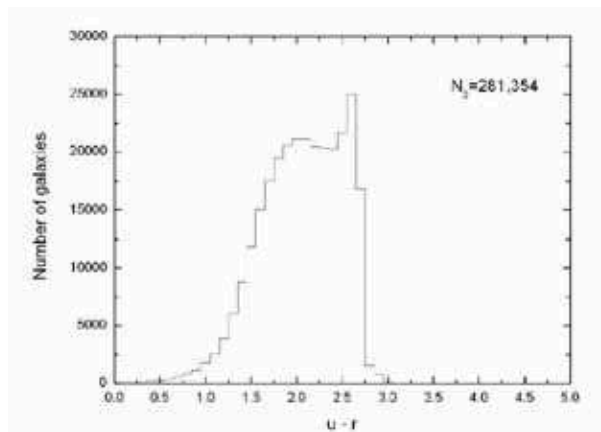}
   \end{center}
   \caption
   { \label{fig:example}
The $u-r$ histogram for late-type galaxies.}
   \end{figure}

     \begin{figure}
   \begin{center}
   \includegraphics[bb=32 179 346 402,width=8cm,clip]{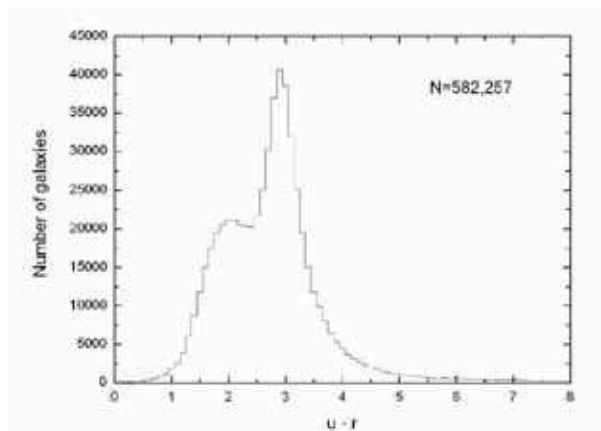}
   \end{center}
   \caption
   { \label{fig:example}
The $u-r$ histogram for the total galaxy samples.}
   \end{figure}

\begin{figure}
\includegraphics[bb=35 172 326 401,width=8cm,clip]{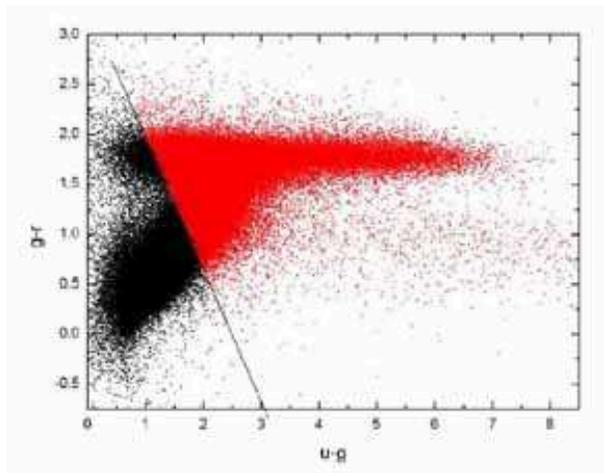}
\caption[]{The $g-r$ versus $u-r$ diagram, red points represent
early-type galaxies and black points represent late-type galaxies.}
 \label{fig6}
\end{figure}

\section{Redshift Estimation}

\subsection{The Used Sample}
Before the experiment of photometric redshift prediction, we
selected the objects from the total galaxy sample satisfying the
following criteria (also see Vanzella et~al. 2004). The obtained
galaxy sample consists of galaxies with $r$-band Petrosian magnitude
brighter than 17.77; the spectroscopic redshift confidence must be
greater than 0.95 and there must be no warning flags. According to
the restriction, we obtained the sample containing 375,929 galaxies
from 582,257 data sets described in Section 4. Whereas 191,200 of
the sample belong to early-type galaxies and 184,729 to late-type
galaxies. The Galactic absorption in the different filters was
obtained from the dust maps of Schlegel et~al. (1998).

\subsection{Result}

ANNs are used to predict photometric redshifts for the selected
galaxy samples. From the total sample, we randomly selected 150,000
for training, 50,000 for validation and the rest 175,929 as test
sample. By training, the regressor is obtained, then it can be used
to predict photometric redshifts of the test sample. The
root-mean-square (rms) redshift error is represented as $\sigma_{\rm
z}$.

Similarly, for the early-type galaxy sample, we randomly partitioned
them into 80,000 for training, 20,000 for validation and 91,200 for
testing, respectively. The late-type galaxy sample was also
separated into training, validation and test sets with respective
sizes 80,000, 20,000 and 84,729. We also applied ANNs to predict the
photometric redshifts of the two subclasses, respectively. Then
calculating their mixed accuracy is as follows:
\begin{equation}
\sigma_{\rm mix}=\sqrt{{\sigma_{\rm early}}^2+{\sigma_{\rm
late}}^2}
\end{equation}

\begin{equation}
\sigma_{\rm
early}=\sqrt{\frac{1}{N_1}({{\sum_{i=1}^{N_1}(\emph{z}NN_i-\emph{z}spec_i)^2}}})
\end{equation}

\begin{equation}
\sigma_{\rm late}=\sqrt{\frac{1}{N_2
}({{\sum_{i=1}^{N_2}(\emph{z}NN_i-\emph{z}spec_i)^2}}})
\end{equation}
where $\emph{z}NN_i$ is the neural output, $\emph{z}spec_i$ is the
target, $N_1$ is the test sample number of early-type galaxies, and
$N_2$ is the test sample number of late-type galaxies. Finally, we
compared the mixed accuracy with that of the total galaxy sample
alone.

The experimental results with different input parameters and
different samples by means of different ANN structures are given in
Table 1. For different samples, the applied ANN structures are
different. When applying ANNs on the actual photometric target
sample, the whole procedure should be run several times with the
test set by modifying the parameters of training (e.g., weight
decay, the number of hidden layers) in order to optimize the
performance. As shown in Table 1., it is evident that the results
based on model magnitudes are better than those based on Petrosian
magnitudes, while those based on dereddened magnitudes are superior
to those based on model magnitudes. We also find that the
performance improves when adding the parameters ($PetR50$,
$PetR90$). For various situations, the rms scatter of photometric
redshifts for early-type galaxies shows better performance than that
for late-type galaxies. Comparing with the rms scatter of the total
galaxy sample, the mixed scatter of photometric redshift estimation
improves when dividing galaxies into early-type ones and late-type
ones. This conclusion is similar to that of Wang et~al. (2007).
Especially for early-type galaxies, the result is rather better than
that of late-type galaxies. For example, when taking dereddened
$u,g,r,i,z,PetR50,PetR90$ as input pattern, the rms deviation of
photometric redshifts for early-type galaxies adds up to 0.0164,
that for late-type galaxies is 0.0217, that for the mixed sample
amounts to 0.0192, and that for the total sample is 0.0196. In order
to see the results clearly, the comparisons of predicted photometric
reshifts with spectroscopic redshifts for early-type galaxies,
late-type galaxies and the overall galaxies are plotted in Figs~6-8.
These figures indicate that the experimental results for early-type
and the overall galaxies show very well performance, although the
result of late-type galaxies is not good. The correlation
coefficient $R$ further proves the conclusion, $R$ is 0.0952 for
early-type galaxies, 0.0881 for late-type galaxies and 0.934 for the
overall galaxies. That early-type galaxies have more better accuracy
of photometric redshifts than late-type ones may be due to the fact
that the spectra of early-type galaxies show a more prominent break
at 4000$\AA$ and therefore a better photo-z signal.

\begin{table*}
\begin{center}
\caption{Photometric redshift prediction with artificial neural
networks}
\bigskip
\begin{tabular}{lccccccl}
\hline\hline      &   Galaxy      &   ANN &for Individual Subset   &for Mixed Sample    &for Total Sample   \\
               \raisebox{2.3ex}[]{Parameters}       &   type                &
             structure
             &$\sigma_{\rm early}$ or $\sigma_{\rm late}$  &$\sigma_{\rm mix}$  &$\sigma_{\rm z}$\\

\hline

     & early-type    &5:10:1    &0.0221     & & \\
 \cline{2-4}   \raisebox{3ex}[]{Petrosian u, g, r, i, z}  &late-type      & 5:5:5:1  &0.0286
& \raisebox{2.3ex}[]{ 0.0254}  & \raisebox{2.3ex}[] {0.0264} \\

\hline
 Petrosian u, g,  r, i, z,     &early-type  &7:5:10:1    &0.0208    &  &  \\
\cline{2-4} PetR50, PetR90        &late-type   &7:5:10:1 &0.0271 & \raisebox{2.3ex}[]{0.024} & \raisebox{2.3ex}[]{0.0247} \\

\hline
     & early-type    &5:10:1    &0.0178     & & \\
 \cline{2-4}   \raisebox{3ex}[]{Model u, g, r, i, z}  &late-type      & 5:10:1
 &0.0230
& \raisebox{2.3ex}[]{ 0.0204}  & \raisebox{2.3ex}[] {0.0214} \\

\hline
 Model u, g,  r, i, z,     &early-type  &7:5:10:1    &0.0166   &  &  \\
 \cline{2-4} PetR50, PetR90        &late-type   &7:5:10:1 &0.0217 & \raisebox{2.3ex}[]{0.0192} & \raisebox{2.3ex}[]{0.0203} \\

\hline
    & early-type    &5:10:1    &0.0174    & & \\
 \cline{2-4}  \raisebox{2.3ex}[] {Dereddened  u, g, r, i, z}  &late-type      & 5:10:1
 &0.0230
& \raisebox{2.3ex}[]{ 0.0203}  & \raisebox{2.3ex}[] {0.0211} \\

\hline
 Dereddened u, g,  r, i, z,     &early-type  &7:5:10:1    &0.0164   &  &  \\
\cline{2-4} PetR50, PetR90        &late-type   &7:5:10:1 &0.0217 & \raisebox{2.3ex}[]{0.0192} & \raisebox{2.3ex}[]{0.0196} \\

\hline

\end{tabular}
\bigskip
\end{center}
\end{table*}

\begin{figure}
\includegraphics[bb=0 1 376 386,width=8cm,clip]{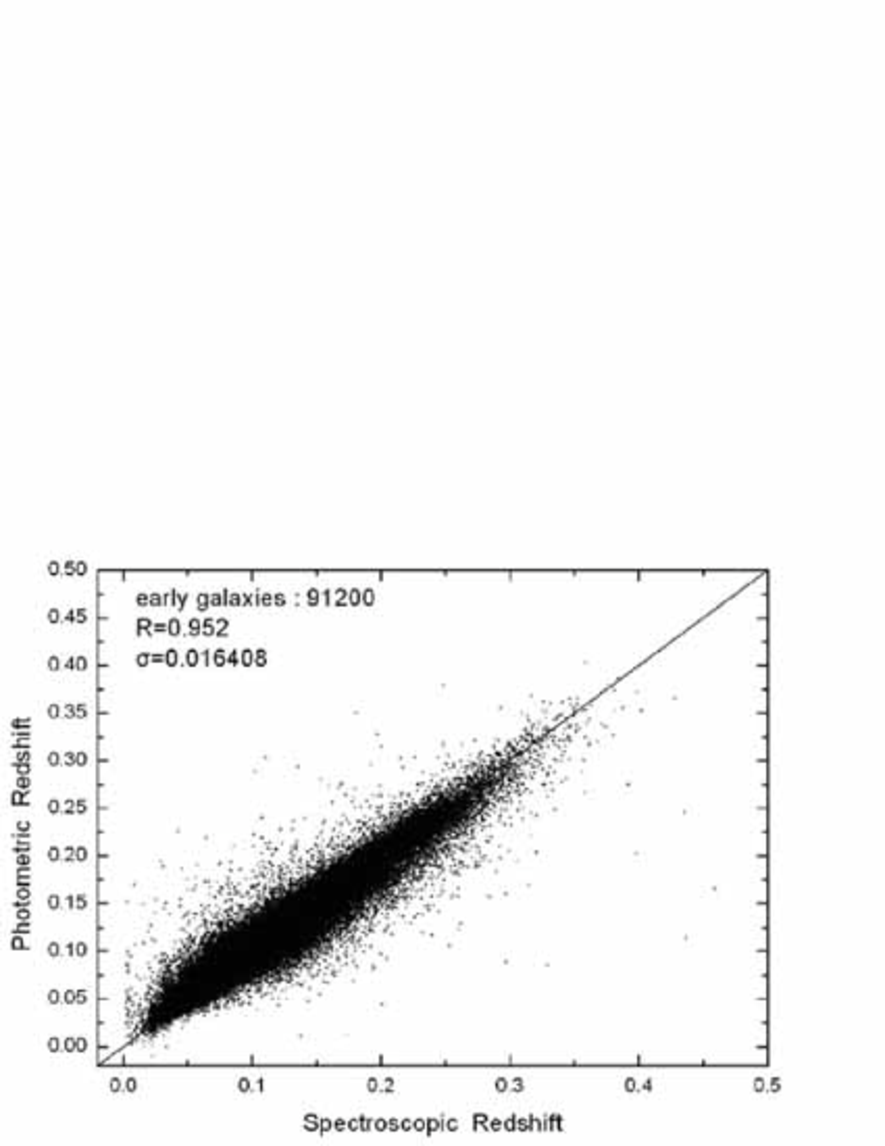}
  \caption{The comparison of predicted photometric reshifts with spectroscopic redshifts for early-type galaxies.}
\end{figure}
\begin{figure}
\includegraphics[bb=12 5 403 299,width=8cm,clip]{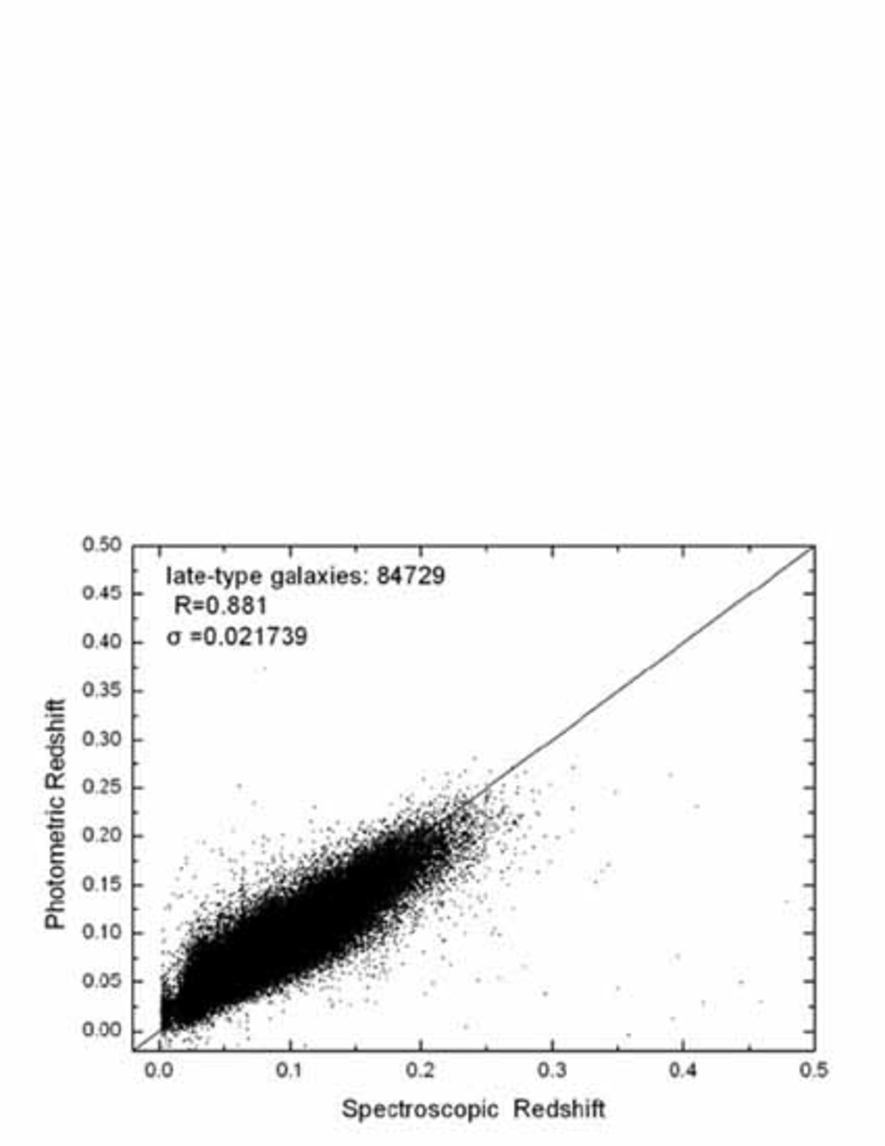}
  \caption{The comparison of predicted photometric reshifts with spectroscopic redshifts for late-type galaxies.}
\end{figure}

\begin{figure}
\includegraphics[bb=12 5 402 302,width=8cm,clip]{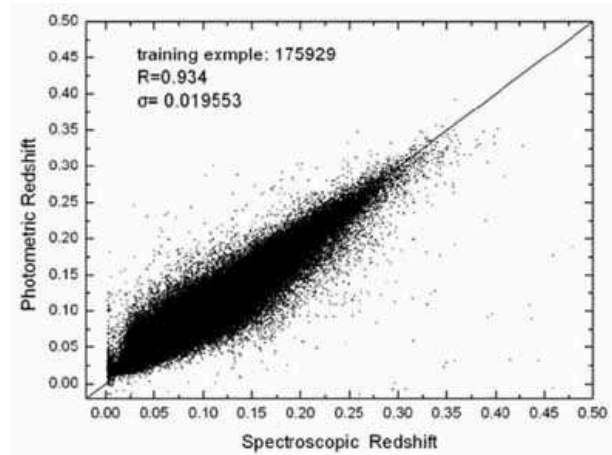}
  \caption{The comparison of predicted photometric reshifts with spectroscopic redshifts for the overall galaxies}
\end{figure}

\section{Conclusions and Discussions}
Firstly we employed an unsupervised method, k-means algorithm to
subdivide the total galaxy sample into two subclasses. The objects
are clustered by its intrinsic properties. By consulting the NED
database, we know that the two subclasses belong to early-type
galaxies and late-type galaxies, respectively. Based on the total
sample and the two subsamples, we have made various experiments with
different parameters for photometric redshift estimation by means of
ANNs. Experimental results indicate that no matter employing
Petrosian magnitudes, model magnitudes or dereddened magnitudes, the
more parameters are considered, the higher the accuracy is. When the
parameters are added in the training data, there will be more
information for the network to improve its capability of prediction
and generalization, so the final accuracy also improves
correspondingly. This result is consistent with the work (Li et~al.
2007). This is a typical characteristics of ANNs, which can be
trained directly on problems with hundreds or thousands of inputs.
Whereas for kernel regression and support vector machines (SVMs),
the optimal choice of input pattern is necessary (Wang et~al. 2007,
2008). Table~1 shows that the accuracy of photometric redshifts with
the mixed sample outperforms that with the overall sample in
different situations. The best experimental result is that the
prediction accuracy is $\sigma_{\rm z}=0.0196$ for the overall
sample, $\sigma_{\rm mix}=0.0192$ for the mixed sample, $\sigma_{\rm
early}=0.0164$ for early-type galaxies, $\sigma_{\rm late}=0.0217$
for late-type galaxies.

Up to now, there are many efforts on photometric redshifts with
ANNs. Because different work is based on different samples,
different attributes and different architectures, we only give a
rough comparison of ANNs which are applied for photometric redshift
estimation in different references, as shown in Table~2. Comparing
with the former work (see Table 2), the scheme that we estimate
photometric redshifts after classifying galaxies into early-type
ones and late-type ones is applicable and satisfactory, moreover the
scheme helps to study galaxies in detail and improve the efficiency
of photometric redshift estimation. The improvement in accuracy of
photometric redshift estimation is of great importance to the study
of large-scale structure of the universe as well as the formation
and evolution of galaxies. When the quality and quantity of
observational data increases, more and more parameters are available
to this problem. Moreover, ANNs will show its superiority in
tackling this complex situation and have wide application (i.e.
classification, regression, feature selection) in astronomy. In
addition, unsupervised approaches don't require human to have the
foreknowledge of the classes, and mainly using some clustering
algorithm to classify data. These procedures can be used to
determine the number and location of the unimodal classes and
helpful for astronomers to find unusual or unknown objects or
phenomenon.

\begin{table*}[h!!!]
\begin{center}
\caption{The comparison of different work for predicting photometric
redshifts using ANNs}
\bigskip
\begin{tabular}{rllllll}
\hline\hline
References &Method    &   Sample   &       $\sigma $\\
\hline
Firth et~al. (2003)      &ANNs    &EDR  &0.0230\\
Collister \& Lahav (2004)&ANNz    &EDR  &0.0229\\
Vanzella et~al. (2004)    &MLP     &DR1  &0.0220\\
Li et~al. (2007)         &MLP      &DR2&0.0202\\
D'Abrusco et~al. (2007)   &MLP      &DR5&0.0208\\
this work                &ANNs      &DR5&0.0192\\

\hline

\end{tabular}
\bigskip
\end{center}
\end{table*}

\section{ACKNOWLEDGMENTS}
We are very grateful to referee's constructive and insightful
suggestions as well as helping us improve writing. This paper is
funded by National Natural Science Foundation of China under Grant
Nos. 10473013, 90412016 and 10778724, and by Chinese National 863
project No.2006AA01A120. This research has made use of data products
from the SDSS survey and from the Two Micron All Sky Survey (2MASS).
The SDSS is managed by the Astrophysical Research Consortium for the
Participating Institutions. The Participating Institutions are the
American Museum of Natural History, Astrophysical Institute Potsdam,
University of Basel, University of Cambridge, Case Western Reserve
University, University of Chicago, Drexel University, Fermilab, the
Institute for Advanced Study, the Japan Participation Group, Johns
Hopkins University, the Joint Institute for Nuclear Astrophysics,
the Kavli Institute for Particle Astrophysics and Cosmology, the
Korean Scientist Group, the Chinese Academy of Sciences (LAMOST),
Los Alamos National Laboratory, the Max-Planck-Institute for
Astronomy (MPIA), the Max-Planck-Institute for Astrophysics (MPA),
New Mexico State University, Ohio State University, University of
Pittsburgh, University of Portsmouth, Princeton University, the
United States Naval Observatory, and the University of Washington.
2MASS is a joint project of the University of Massachusetts and the
Infrared Processing and Analysis Center/California Institute of
Technology, funded by the National Aeronautics and Space
Administration and the National Science Foundation. This research
has made use of the NASA/IPAC Extragalactic Database (NED) which is
operated by the Jet Propulsion Laboratory, California Institute of
Technology, under contract with the National Aeronautics and Space
Administration. This research has also made use of the SIMBAD
database, operated at CDS, Strasbourg, France.


\begin{thebibliography}{}
\bibitem {}
Bailer-Jones C.A.L., Gupta R., Singh H.P., 2001, Automated Data
Analysis in Astronomy, R. Gupta, H.P. Singh, C.A.L. Bailer-Jones
(eds.), Narosa Publishing House, New Delhi, India, 51-68

\bibitem {}
Bailer-Jones C.A.L., 1996, PhD thesis, Univ. Cambridge

\bibitem {}
Bailer-Jones C.A.L., 2000, A\&A, 357, 197

\bibitem {}
Ball N. M., Brunner R. J., Myers A. D., Strand N. E., Alberts S. L.,
Tcheng D., Lior$\acute{a}$X., 2007, ApJ, 663, 774

\bibitem {}
Ball N. M., Loveday J., Fukugita M., et~al. 2004, MNRAS, 3, 348

Baum W. A., 1962, in George C. M., ed, Proc. IAU Symp. 15,
Photoelectric Magnitudes and Red-Shifts. Macmillan Press, New York,
NY, p. 390

\bibitem {}
Bertin E., Arnouts S., 1996, A\&AS, 117, 393

\bibitem {}
Bishop C. M., 1995, Neural Networks for Pattern Recognition, Oxford:
Oxford University Press

\bibitem {}
Brunner R. J., Connolly A. J., Szalay A. S., Bershady M. A., 1997,
ApJ, 482, L21

\bibitem {}
Bruzual A. G., Charlot S., 1993, ApJ, 405, 538

\bibitem {}
Budav$\acute{a}$ri T. et~al., 2005, ApJ, 619, L31

\bibitem {}
Coleman G. D., Wu C. C., Weedman D. W., 1980, ApJS, 43, 393

\bibitem {}
Collister A. A., Lahav O., 2004, PASP, 116, 345

\bibitem {}
Connolly A. J., Csabai I., Szalay A. S., Koo D. C., Kron R. G., Munn
J. A., 1995, AJ, 110, 2655

\bibitem {}
Couch W. J., Ellis R. S., Godwin J., Carter D., 1983, MNRAS, 205,
1287

\bibitem {}
Csabai I. et al., 2003, ApJ, 125, 580

\bibitem {}
Curtis H.D. 1918, Pub. Lick Obs. 13, 55

\bibitem {}
D'Abrusco R., Staiano A., Longo G., Brescia M., Paolillo M., De
Filippis E., Tagliaferri R., 2007, ApJ, 663, 752

\bibitem {}
Firth A. E., Lahav, O., Somerville, R. S. 2003, MNRAS, 4, 339

\bibitem {}
Firth A. E., Lahav O., Somerville R. S., 2003, MNRAS, 339, 1195

\bibitem {}
Folkes S. R., Lahav O., Maddox S. J., 1996, MNRAS, 283, 651

\bibitem {}
Gao D., Zhang Y., Zhao Y., 2008, in Robert W. Argyle, Peter S.
Bunclark, and James R. Lewis., eds, ASP Conference Series, Vol.394,
Astronomical Data Analysis Software and Systems, Kensington Town
Hall, London, United Kingdom, p.525

\bibitem {}
Haykin S., Neural Networks: A Comprehensive Foundation, Macmillan
College Publishing Company, New York, 1994

\bibitem {}
Hsieh B. C., Yee H. K. C., Lin H., Gladders M. D., 2005, ApJS, 158,
161

\bibitem {}
Hubble E. Humason M.L. 1931, ApJ, 74, 43

\bibitem {}
Levenberg K., 1944, Quarterly Journal of Applied MathematicII, 2,
164

\bibitem {}
Koo D. C., 1985, AJ, 90, 418

\bibitem {}
Lahav O., Naim A., Sodr$\acute{e}$ L. J., Storrie-Lombardi M. C.,
1996, MNRAS, 283, 207

\bibitem {}
Li L., Zhang Y., Zhao Y., Yang D., 2006, Progress in Astronomy,
24(4), 285

\bibitem {}
Li L., Zhang Y., Zhao Y., Yang D., 2007, ChJAA, 7, 448

\bibitem {}
MacQueen J. B., 1967, Proceedings of 5-th Berkeley Symposium on
Mathematical Statistics and Probability, Berkeley, University of
California Press, 1:281-297

\bibitem {}
Marquardt D. W., 1963, Journal of the Society of Industrial and
Applied Mathematics, 11, 431

\bibitem {}
Melnick J., Sargent W.L.W. 1977, ApJ, 215, 401

\bibitem {}
Miller, A. S.£¬1993, Vistas in Astronomy£¬36(2), 141

\bibitem {}
Nadaraya E. A., 1964, Theory of Probability and its Applications, 9,
141

\bibitem {}
Nielsen M. L., Odewahn S. C., 185th AAS Meeting, 1994, 26, 1498

\bibitem {}
Odewahn S. C., Nielsen M.L., 1994, Vistas in Astronomy, 3, 38

\bibitem {}
Oemler A. Jr. 1974, ApJ 194, 1

\bibitem {}
Serra-Ricart M., Calbet X., Garrido L., Gaitan V., 1993, AJ, 106,
1685

\bibitem {}
Schlegel D. J., Finkbeiner D. P., Davis M., 1998, ApJ, 500, 525

\bibitem {}
Shimasaku, K., et al. 2001, AJ, 122, 1238

\bibitem {}
Storrie-Lombardi M. C., Lahav O., 1994, guest eds, Vistas in
Astron., spec. iss. on ANNs in Astronomy, 38 (3)

\bibitem {}
Strateva I. et al., 2001, AJ, 122, 1861

\bibitem {}
Suchkov A. A., Hanisch R. J., Margon B., 2005, AJ, 130, 2439

\bibitem {}
Vanzella E. et al., 2004, A\&A, 423, 761

\bibitem {}
Wadadekar Y., 2005, PASP, 117, 79

\bibitem {}
Wang D., Zhang Y., Liu C., Zhao Y., 2007, MNRAS, 382, 1601

\bibitem {}
Wang D., Zhang Y., Liu C., Zhao Y., 2008, ChJAA, 8(1), 119

\bibitem {}
Wang Y., Bahcall N., Turner E. L., 1998, AJ, 116, 2081

\bibitem {}
Way M. J., Srivastava A. N., 2006, ApJ, 647, 102

\bibitem {}
Weaver W. B., 2000, ApJ, 1, 541

\bibitem {}
York D. G. et al., 2000, AJ, 120, 1579

\bibitem {}
Zhang Y., Zhao Y., 2004, A\&A, 422, 1113

\bibitem {}
Zhang Y., Zhao Y., 2007, ChJAA, 7, 289

\end{thebibliography}
\end{document}